\documentclass[aps,floats]{revtex4}
\usepackage{amsmath,amssymb}
\usepackage{graphicx,epsfig}
\usepackage[greek,english]{babel}
\usepackage{bbold}

\begin{document}
\bibliographystyle {plain}

\def\oppropto{\mathop{\propto}} 
\def\opsimeq{\mathop{\simeq}}
\def\opoverderline{\mathop{\overline}}
\def\operarrow{\mathop{\longrightarrow}}
\def\opsim{\mathop{\sim}}

\def\fig#1#2{\includegraphics[height=#1]{#2}}
\def\figx#1#2{\includegraphics[width=#1]{#2}}


\title{ Even and odd normalized zero modes in random interacting Majorana models  \\
respecting the Parity $P$ and the Time-Reversal-Symmetry $T$  } 


\author{ C\'ecile Monthus }
 \affiliation{Institut de Physique Th\'{e}orique, 
Universit\'e Paris Saclay, CNRS, CEA,
91191 Gif-sur-Yvette, France}

\begin{abstract}
For random interacting Majorana models where the only symmetries are the Parity $P$ and the Time-Reversal-Symmetry $T$, various approaches are compared to construct exact even and odd normalized zero modes $\Gamma$ in finite size, i.e. hermitian operators that commute with the Hamiltonian, that square to the Identity, and that commute (even) or anticommute (odd) with the Parity $P$. Even Normalized Zero-Modes $\Gamma^{even}$ are well known under the name of 'pseudo-spins' $\tau^z_n$ in the field of Many-Body-Localization or more precisely 'Local Integrals of Motion' (LIOMs) in the Many-Body-Localized-Phase where the pseudo-spins happens to be spatially localized. Odd Normalized Zero-Modes $\Gamma^{odd}$ are popular under the name of 'Majorana Zero Modes' or 'Strong Zero Modes'. Explicit examples for small systems are described in detail. Applications to real-space renormalization procedures based on blocks containing an odd number of Majorana fermions are also discussed.

\end{abstract}

\maketitle

\section{ Introduction   }

In the field of quantum interacting models, the notion of Normalized Zero Modes has emerged as an essential idea in various contexts recently.
Here, a zero-mode $\Gamma$ will be defined as an hermitian operator
\begin{eqnarray}
\Gamma^{\dagger} = \Gamma
\label{gahermitian}
\end{eqnarray}
that commutes with the Hamiltonian 
\begin{eqnarray}
[H, \Gamma ]=0
\label{gazero}
\end{eqnarray}
This zero-mode will be called Normalized if it squares to the Identity
\begin{eqnarray}
\Gamma^2 =  \mathbb{1}
\label{gasquareunity}
\end{eqnarray}
In addition, in models with a Parity operator $P$ that commutes with the Hamiltonian $H$ (see Eq. \ref{paritytotal} below for Majorana models)
\begin{eqnarray}
[H,P]   = 0
\label{hp}
\end{eqnarray}
the normalized Zero Mode $\Gamma$ will be called Even if it commutes with the Parity $P$ 
\begin{eqnarray}
[\Gamma^{even} ,P]=0
\label{gaeven}
\end{eqnarray}
or Odd if it anticommutes with the parity $P$ 
\begin{eqnarray}
\{ \Gamma^{odd} ,P\}=0
\label{gaodd}
\end{eqnarray}

Even Normalized Zero-Modes $\Gamma^{even}$ have become very popular recently under the name of pseudo-spins $\tau^z_n$ that commute with each other
 and with the Hamiltonian $H$ in the field of Many-Body-Localization (see the recent reviews \cite{revue_huse,revue_altman,revue_vasseur,revue_imbrie,revue_rademaker,review_mblergo,review_prelovsek,review_rare} and references therein) :
in the Many-Body-Localized-Phase, these pseudo-spins are spatially localized and are then called 
 Local Integrals of Motion (LIOMs) \cite{emergent_swingle,emergent_serbyn,emergent_huse,emergent_ent,imbrie,serbyn_quench,emergent_vidal,emergent_ros,
emergent_rademaker,serbyn_powerlawent,c_emergent,ros_remanent,wortis,c_liomsper,counting_lioms}.

Odd Normalized Zero-Modes $\Gamma^{odd}$ have also attracted a lot of interest recently under the name of Majorana Zero Modes (MZM)
in the context of the classification of topological phases \cite{kitaevchain,kitaevfid,kitaevalpha,10phases}.
They have been considered both in random systems in relation with Many-Body-Localization models \cite{maj_pol}
or in non-random models like the integrable XYZ chain \cite{strongzeromode}
where they were called 'Strong Zero Mode',
 with various consequences for the long coherence time of edge spins 
\cite{longcoherence,mila} and the phenomenon of prethermalization \cite{prethermal}.
It should be stressed that in the present work, the commutator with $H$ is required to be exactly zero in finite size (Eq. \ref{gazero}), instead of being exponentially small in the system size as defined in \cite{strongzeromode}.
The present definition will be more convenient technically for our present purposes, and can always be achieved by some appropriate choice
of the boundary couplings. As explained in more details in section \ref{sec_pairing},  
one can for instance put to zero the couplings involving the last Majorana operator $\gamma_{2N}$,
so that the modified Hamiltonian involving an odd number $(2N-1)$ of Majorana operators has exact odd zero modes 
\cite{akhmerov,goldstein,wilczek,feldman,moreabout,kauffman}.

In the present work, the goal is to compare various approaches to construct these even and odd normalized zero-modes
in random interacting Majorana models, where the only symmetries are the Parity $P$ and the Time-Reversal-Symmetry $T$.
The term 'random' here means that the couplings are arbitrary, so that the construction is completely general
and neither translation invariance nor quantum integrability are required (even if further simplifications are to be expected for integrable models as discussed in the conclusion of Ref \cite{strongzeromode}).

The paper is organized as follows.
In Section \ref{sec_models}, the notations are introduced for Majorana models with special boundary conditions
to insure the existence of odd zero modes in finite size.
The Even and Odd normalized zero modes are discussed respectively in sections  \ref{sec_even} and \ref{sec_odd}.
In section \ref{sec_dyn}, the matrix describing the dynamics within the subspace of odd operators of Ref \cite{goldstein}
is adapted to take into account the presence of the Time-Reversal-Symmetry $T$, where the odd operators can be classified
 with respect to the two flavors $T=\pm 1$.
To see how this general formalism works in practice, the cases where the Hamiltonian depends only on $2N-1=3$ and $2N-1=5$
are described in detail in sections \ref{sec_three} and \ref{sec_five} respectively.
Our conclusions are summarized in section \ref{sec_conclusion}.
The Appendix \ref{sec_app} contains the translation of various notions in the quantum spin chain language.

\section{ Majorana models respecting $P$ and $T$ with special boundary conditions }

\label{sec_models}

\subsection{ Models involving $2N$ Majorana fermions }

Many interacting quantum models with an Hilbert space of size $2^N$, involving either $N$ quantum spins $S=1/2$ (see the Appendix) or $N$ spinless Dirac fermions,
can be reformulated in terms of the even number $2N$ of Majorana operators $\gamma_j$ with $j=1,..,2N$,
that are hermitian
\begin{eqnarray}
\gamma_j^{\dagger}=\gamma_j
\label{hermi}
\end{eqnarray}
 square to the Identity
\begin{eqnarray}
\gamma_j^2=  \mathbb{1}
\label{squareunity}
\end{eqnarray}
and anti-commute with each other
\begin{eqnarray}
\{ \gamma_j , \gamma_l \} \equiv \gamma_j \gamma_l + \gamma_l \gamma_j && = 0 \ \ \ \ \ \  { \rm for } \ \ \   j \ne l
\label{anticomm}
\end{eqnarray}
We will be interested in models where the Hamiltonian $H$ commutes
\begin{eqnarray}
[H,P]=0
\label{commhp}
\end{eqnarray}
with the total parity 
\begin{eqnarray}
P   = i^N \gamma_1 \gamma_2 \gamma_{3} \gamma_{4}   ... \gamma_{2N-1} \gamma_{2N}
\label{paritytotal}
\end{eqnarray}
i.e. the Hamiltonian can only contain interactions between an even number of Majorana fermions,
like two-Majorana, four-Majorana, six-Majorana, etc.

\subsection{Time-Reversal-Symmetry $T$  }

Another possible very common symmetry is the Time-Reversal-Symmetry $T$, which is an anti-unitary symmetry so that it
is simpler to define it via its action on $i$ and on the elementary Majorana operators 
\cite{kitaevchain,kitaevfid,kitaevalpha,10phases}
\begin{eqnarray}
T i T^{-1} && =  -i
\nonumber \\
T \gamma_{2j-1} T^{-1} &&=  \gamma_{2j-1}
\nonumber \\
T \gamma_{2j} T^{-1} && = - \gamma_{2j}
\label{time}
\end{eqnarray}
It is then useful to relabel the Majorana operators with the flavors $a$ and $b$ to stress their different behaviors with respect to $T$
\begin{eqnarray}
 \gamma_{2j-1} && = a_j
\nonumber \\
 \gamma_{2j} && = b_j
\label{timeab}
\end{eqnarray}

\subsection{  Boundary conditions producing an exact pairing in the spectrum } 

\label{sec_pairing}

In this paper, we will focus on the case where the Hamiltonian for the $(2N)$ Majorana fermions $(\gamma_1,...,\gamma_{2N})$
actually does not involve the last one $\gamma_{2N}$, but only involves the odd number $(2N-1)$ of Majorana fermions $(\gamma_1,...,\gamma_{2N-1})$,
a problem that has attracted a lot of interest recently
\cite{akhmerov,goldstein,wilczek,feldman,moreabout,kauffman}.
Then the Hamiltonian $H$ commutes both with $\gamma_{2N}$
\begin{eqnarray}
 [H , \gamma_{2N}]=0
\label{hgn}
\end{eqnarray}
and with the Parity of Eq. \ref{paritytotal} that can be rewritten as
\begin{eqnarray}
P   = i {\Upsilon}^{tot}  \gamma_{2N}
\label{parityups}
\end{eqnarray}
in terms of the hermitian odd operator \cite{akhmerov,goldstein,wilczek,feldman,moreabout,kauffman}
\begin{eqnarray}
{\Upsilon}^{tot} \equiv    -i P \gamma_{2N} =  i^{N-1}   \gamma_{1} \gamma_{2} ... \gamma_{2N-2}\gamma_{2N-1} 
\label{upstot}
\end{eqnarray}
that squares to the Identity
\begin{eqnarray}
({\Upsilon}^{tot})^2 =  \mathbb{1}
\label{ugasquareunity}
\end{eqnarray}
and that commutes with $H$
\begin{eqnarray}
[ H, {\Upsilon}^{tot} ] =0
\label{hupstot}
\end{eqnarray}
This operator ${\Upsilon}^{tot} $ thus satisfies all the properties of an odd normalized zero mode (Eqs 
\ref{gahermitian} \ref{gazero} \ref{gasquareunity} \ref{gaodd}).

From the point of view of the spectrum of the Hamiltonian, 
this means that there exists an exact pairing between the eigenstates of the two Parity sectors $P=\pm 1$ \cite{strongzeromode}.
More precisely, the  diagonalization of $H$ in the even sector $P=+1$
involves
\begin{eqnarray}
{\cal N} \equiv 2^{N-1}
\label{caln}
\end{eqnarray}
even eigenstates
\begin{eqnarray}
\vert n^e>=P  \vert n^e>
\label{pketeven}
\end{eqnarray}
 of eigenvalues $ E_{n}  $,
with the spectral decomposition
\begin{eqnarray}
H_{even}  && = \sum_{n=1}^{{\cal N}}  E_{n} \pi_{n^e} 
\label{spectrale}
\end{eqnarray}
in terms of the projectors
\begin{eqnarray}
\pi_{n^e} && = \vert n^e> < n^e \vert
\label{proje}
\end{eqnarray}
Then the state obtained by the application of the operator of Eq. \ref{upstot}
\begin{eqnarray}
\vert n^o> && = {\Upsilon}^{tot} \vert n^e>
\label{none}
\end{eqnarray}
belongs to the Parity sector $P=-1$ as a consequence of the anticommutation ${\Upsilon}^{tot} P= -P {\Upsilon}^{tot}  $
\begin{eqnarray}
P \vert n^o> && =P  {\Upsilon}^{tot}  \vert n^e> = -  {\Upsilon}^{tot}  P \vert n^e> = -  {\Upsilon}^{tot}   \vert n^e>= - \vert n^o>
\label{pno}
\end{eqnarray}
while it is an eigenstate of $H$ with the same eigenvalue $E_n$ as a consequence of the commutation ${\Upsilon}^{tot} H= H {\Upsilon}^{tot}  $
\begin{eqnarray}
H \vert n^o> && = H   {\Upsilon}^{tot}   \vert n^e>=     {\Upsilon}^{tot}  H \vert n^e> = E_n   {\Upsilon}^{tot}   \vert n^e>= E_n   \vert n^o>
\label{hno}
\end{eqnarray}
So the spectral decomposition in the odd sector $P=-1$ reads
\begin{eqnarray}
H_{odd}  && = \sum_{n=1}^{{\cal N}}  E_{n} \pi_{n^o} 
\label{spectral}
\end{eqnarray}
in terms of the projectors 
\begin{eqnarray}
\pi_{n^o} && = \vert n^o> < n^o \vert
\label{projo}
\end{eqnarray}
So the ${\cal N} $  energy levels $E_n$ are all twice degenerated, 
and the two corresponding eigenstates $\vert n^e> $ and $\vert n^o> $ belong to the two Parity sectors $P=\pm 1$.

\section{ Even zero modes }

\label{sec_even}

\subsection{ Subspace generated by the $2^N$ orthogonal projectors onto eigenstates } 

Since the $2^N= 2 {\cal N}$ orthogonal projectors $\pi_{n^e}  $ and $\pi_{n^o}  $
are hermitian operators that
commute with the Hamiltonian $H$ and with the Parity $P$,
 any linear combination with real coefficients $(c_{n^e},c_{n^o}) $ of them produces an even zero mode
\begin{eqnarray}
Z^{even}  =  \sum_{n=1}^{{\cal N}} ( c_{n^e} \pi_{n^e} +  c_{n^o} \pi_{n^o} )
\label{ze}
\end{eqnarray}

\subsection{ Subspace generated by  the first ${\cal N}=2^{N-1}$ powers of the Hamiltonian } 

The expansion onto orthogonal projectors associated to eigenstates requires the diagonalization of the Hamiltonian.
Yang and Feldman \cite{feldman} have thus proposed to construct instead
${\cal N}=2^{N-1}$ even zero modes directly from 
 the first powers of the Hamiltonian \cite{feldman}
\begin{eqnarray}
Z^{even}_p && = H^p \ \ \ {\rm with } \ \ p=0,1,..,{\cal N}-1
\label{evenpower}
\end{eqnarray}

The link with the expansion onto the orthonormal projectors of Eq. \ref{ze} reads
\begin{eqnarray}
Z^{even}_p && =  \sum_{n=1}^{{\cal N}} E_n^k ( \pi_{n^e} +\pi_{n^o} ) =H_{even}^p + H_{odd}^p
\label{evenzeromodepi}
\end{eqnarray}

Note that the next power corresponding to $p={\cal N}$ is not independent of the previous ones as a consequence of the Cayley-Hamilton theorem 
in each parity sector. Indeed in the even sector,
the ${\cal N}$ eigenvalues $E_{n}$ are the solutions of the characteristic polynomial of degree ${\cal N}$
\begin{eqnarray}
0= det( E - H_{even})  && = \prod_{n=1}^{\cal N} (E-E_{n}) 
= E^{ \cal N} + c_1 E^{ {\cal N } -1} + c_2 E^{ {\cal N } -2} ...   +c_{ {\cal N}-1} E + c_{\cal N} 
\label{cayleyH}
\end{eqnarray}
So the power $H_{even}^{\cal N}  $ is given by the same linear combination in terms of the previous powers of $H_{even} $
\begin{eqnarray}
 H_{even}^{\cal N} = \sum_{n=1}^{{\cal N}} E_{n}^{\cal N} \pi_{n^e} 
 = - c_1 H_{even}^{ {\cal N } -1} - c_2 H_{even}^{ {\cal N } -2} ...   -c_{ {\cal N}-1} H_{even} + c_{\cal N} 
\label{hpowerrec}
\end{eqnarray}
Similarly, within the odd sector, as a consequence of the exact pairing of the spectrum, the same linear combination of the Cayley-Hamilton theorem holds
\begin{eqnarray}
 H_{odd}^{\cal N} = \sum_{n=1}^{{\cal N}} E_{n}^{\cal N} \pi_{n^o} 
 = - c_1 H_{odd}^{ {\cal N } -1} - c_2 H_{odd}^{ {\cal N } -2} ...   -c_{ {\cal N}-1} H_{odd} + c_{\cal N} 
\label{hpowerreco}
\end{eqnarray}
So the full Hamiltonian also satisfies the same equation
\begin{eqnarray}
 H^{\cal N}  && =  H_{even}^{\cal N} + H_{odd}^{\cal N}  = - c_1 H^{ {\cal N } -1} - c_2 H^{ {\cal N } -2} ...   -c_{ {\cal N}-1} H + c_{\cal N} 
\label{cayleyhamilton}
\end{eqnarray}

\subsection{ Diagonalization in terms of pseudo-Majorana fermions }

Via the unitary transformation $U$ that diagonalizes the Hamiltonian $H$,
the Majorana operators $\gamma_j$ with $j=1,2,..,2N-1$ are transformed into the pseudo-Majorana operators
\begin{eqnarray}
\tilde \gamma_j && = U \gamma_j U^{\dagger} 
\label{deftildeg}
\end{eqnarray}
that inherit the anti-commutation relations of the initial Majorana operators (Eq. \ref{anticomm}) and their flavors with respect to $T$ (Eq. \ref{timeab}).
The Hamiltonian can be then rewritten in terms of the $(N-1)$ commuting pseudo-spins operators $j=1,..,N-1$
\begin{eqnarray}
\tau_j^z=i \tilde b_j \tilde a_{j+1}
\label{deftau}
\end{eqnarray}
as the expansion
 \begin{eqnarray}
H &&  = \sum_{p=0}^{N-1} \sum_{1 \leq j_1 <j_2 .. <j_p \leq N-1} \omega^{(p)}_{j_1 j_2...j_p} \tau^z_{j_1} \tau^z_{j_2} ... \tau^z_{j_p} 
\nonumber \\
&& =
\omega^{(0)} +  \sum_{j=1}^{N-1} \omega^{(1)}_j \tau^z_j + \sum_{1 \leq j_1 < j_2 \leq N-1} \omega^{(2)}_{j_1 j_2} \tau^z_{j_1} \tau^z_{j_2}+... 
+  \omega^{(N-1)}_{1,2,..,N-1}  \tau^z_{1} \tau^z_{2}... \tau^z_{N-1}
\label{hmbl}
\end{eqnarray}
where the ${\cal N}=2^{N-1}$ pseudo-couplings $\omega^{(p)}_{j_1 j_2...j_p}  $ allow to reproduce the ${\cal N}=2^{N-1}$ energy levels $E_n$.
The first pseudo-Majorana fermion
\begin{eqnarray}
\tilde \gamma_1 = \tilde a_{1}  
\label{deftilde1}
\end{eqnarray}
is absent from the Hamiltonian of Eq. \ref{hmbl} and is thus an odd normalized zero mode.
Its pairing with the last Majorana fermion $\gamma_{2N}=b_N$ also absent from $H$ produces the last pseudo-spin
\begin{eqnarray}
\tau_N^z = i b_N \tilde a_{1}  
\label{deftaun}
\end{eqnarray}
absent from $H$ that labels the double degeneracy of each energy level $E_n$.

\subsection{ Conclusion on Even Normalized Zero Modes } 

The diagonalization in terms of pseudo-spins in Eq. \ref{hmbl} means that 
all the operators of the form
 \begin{eqnarray}
\Gamma^{e(k)}_{j_1,..,j_k} = \tau^z_{j_1} \tau^z_{j_2}...\tau^z_{j_k}
\label{gammae}
\end{eqnarray}
are normalized even zero modes 
that square to the Identity 
 \begin{eqnarray}
(\Gamma^{e(k)}_{j_1,..,j_k} )^2= \mathbb{1}
\label{gammaecarre}
\end{eqnarray}
These $2^N$ Even Normalized Zero Modes 
can be reconstructed from the knowledge of the $N$ independent pseudo-spins $\tau^z_j $ of Eq. \ref{deftau}.

Since these general definitions of pseudo-Majorana fermions (Eq. \ref{deftildeg}) and pseudo-spins (Eq. \ref{deftau})
may seem somewhat formal and elusive, 
 it is useful to have a more concrete picture in simple cases.
Besides the perturbative expansions discussed for Many-Body-Localized models \cite{ros_remanent,c_liomsper}
and for the edge Majorana localized mode in the integrable pure XYZ chain \cite{strongzeromode},
it is thus instructive to consider small systems to obtain explicit exact results for any values of the couplings :

(i) in Ref \cite{wortis} concerning the two-site Anderson-Hubbard model, the authors have proposed to compute 
first the $16$ many-body-eigenstates and to label them in terms of $4$ pseudo-spins
with the criterion that these $4$ pseudo-spins should be as local as possible (see \cite{wortis} for more details).

(ii) an alternative possibility consists in computing the pseudo-couplings and the pseudo-spins
via the identification of the first powers of the Hamiltonian of Eq. \ref{evenpower},
as described in more details in section \ref{sec_five} B on a specific example.

\section{ Odd Zero Modes } 

\label{sec_odd}

\subsection{ Correspondence between even and odd zero modes via the operator $\Upsilon^{tot}$ } 

Yang and Feldman \cite{feldman} have proposed to use the operator $\Upsilon^{tot}$ of Eq. \ref{upstot} 
in order to transform any even zero mode $Z^{even}$ into 
an odd zero mode by
\begin{eqnarray}
Z^{odd} && = Z^{even} {\Upsilon}^{tot} 
\label{zerooddp}
\end{eqnarray}
It is now interesting to apply this recipe 
to the various even zero modes described in the previous section.

For the projectors of Eqs \ref{proje} and \ref{projo}, one obtains using Eq. \ref{none} the operators
\begin{eqnarray}
\pi_{n^e} {\Upsilon}^{tot}  && = \vert n^e> < n^o \vert
\nonumber \\
\pi_{n^o} {\Upsilon}^{tot}  && = \vert n^o> < n^e \vert
\label{projups}
\end{eqnarray}
that relates the two states of different parities of the same energy level $E_n$.
So the linear combination of Eq. \ref{ze} becomes the linear combinations with real coefficients $(c_n, d_n)$
\begin{eqnarray}
Z^{odd}  =  \sum_{n=1}^{{\cal N}} ( c_{n} \vert n^e> < n^o \vert +  d_n \vert n^o> < n^e \vert )
\label{zo}
\end{eqnarray}

In particular, the even zero modes based on the first ${\cal N}=2^{N-1}$ powers of the Hamiltonian (Eq. \ref{evenpower})
become the odd zero modes
\begin{eqnarray}
Z^{odd}_p && = H^p {\Upsilon}^{tot} = {\Upsilon}^{tot} H^p \ \ \ {\rm with } \ \ p=0,1,..,2^{N-1}-1
\label{oddpower}
\end{eqnarray}

\subsection{ Odd normalized zero modes }

The even normalized zero modes defined in terms of pseudo-spins (Eq. \ref{gammae}) 
become the odd normalized zero modes
 \begin{eqnarray}
\Gamma^{o(k)}_{j_1,..,j_k} = \Gamma^{e(k)}_{j_1,..,j_k}  {\Upsilon}^{tot} =  {\Upsilon}^{tot}  \Gamma^{e(k)}_{j_1,..,j_k} 
\label{gammaoe}
\end{eqnarray}
since they inherit the property to square to the Identity 
 \begin{eqnarray}
(\Gamma^{o(k)}_{j_1,..,j_k} )^2= \mathbb{1}
\label{gammaocarre}
\end{eqnarray}

To see more clearly their physical meaning in terms of the pseudo-spins and pseudo-Majorana operators that diagonalize the Hamiltonian (Eq. \ref{hmbl})
\begin{eqnarray}
\tau_j^z=i \tilde b_j \tilde a_{j+1}
\label{deftaubis}
\end{eqnarray}
with the expansion
 \begin{eqnarray}
H &&   =
\omega^{(0)} +  \sum_{j=1}^{N-1} \omega^{(1)}_j (i \tilde b_j \tilde a_{j+1}) + \sum_{1 \leq j_1 < j_2 \leq N-1} \omega^{(2)}_{j_1 j_2}  (i \tilde b_{j_1} \tilde a_{j_1+1}) (i \tilde b_{j_2} \tilde a_{j_2+1})+... 
+  \omega^{(N-1)}_{1,2,..,N-1}   (i \tilde b_{1} \tilde a_{2}) (i \tilde b_{2} \tilde a_{3}) ... (i \tilde b_{N-1} \tilde a_{N})
\nonumber
\end{eqnarray}
 it is useful to rewrite the operator $\Upsilon^{tot}$ of Eq. \ref{upstot} as
\begin{eqnarray}
 {\Upsilon}^{tot}  =  i^{N-1} \tilde \gamma_1 \tilde \gamma_2    ... \tilde \gamma_{2N-2}  \tilde \gamma_{2N-1}
=  i^{N-1} \tilde a_1 \tilde b_1 \tilde a_2    ... \tilde b_{N-1}  \tilde a_{N}
\label{upstottilde}
\end{eqnarray}

So in the correspondence between even and odd normalized zero modes of Eq. \ref{gammaoe},
the case $k=0$ corresponds to
 \begin{eqnarray}
\Gamma^{e(k=0)} && =\mathbb{1}
\nonumber \\
\Gamma^{o(k=0)} && =  {\Upsilon}^{tot} =\tilde \gamma_1 (i\tilde \gamma_2 \tilde \gamma_3)    ... (i \tilde \gamma_{2N-2} \tilde \gamma_{2N-1} ) 
=   \tilde a_1 (i \tilde b_1 \tilde a_2)    ... (i \tilde b_{N-1}  \tilde a_{N} )
\label{gammaoezero}
\end{eqnarray}
while the case $k=N-1$ corresponds to
 \begin{eqnarray}
\Gamma^{e(k=N-1)}_{1,2,...,N-1} && =(i \tilde\gamma_2 \tilde\gamma_3)
(i \tilde\gamma_4 \tilde\gamma_5)
... (i \tilde\gamma_{2N-2} \tilde\gamma_{2N-1})
\nonumber \\
\Gamma^{o(k=N-1)}_{1,2,..,N-1} && =\tilde \gamma_{1} = \tilde a_{1} 
\label{gammaoelast}
\end{eqnarray}

In conclusion, the odd normalized zero modes are given by the elementary pseudo-Majorana fermion
 $\tilde \gamma_{1} = \tilde a_{1} $ (Eq. \ref{gammaoelast}) absent from the Hamiltonian,
and by the product of $\tilde \gamma_{1} = \tilde a_{1} $ times any number of the pseudo-spins that diagonalize the Hamiltonian,
up to the maximal case given by ${\Upsilon}^{tot}  $ of Eq. \ref{gammaoezero}.

\section{ Dynamics within the subspace of odd operators }

\label{sec_dyn}

In this section, the goal is to adapt the formalism of Ref \cite{goldstein}
to the presence of the Time-Reversal-Symmetry $T$, where the odd operators can be classified
 with the two flavors $T=\pm 1$.

\subsection{ Reminder on the orthonormal basis of the subspace of odd operators }

For an odd number $(2N-1)$ of Majorana fermions $(\gamma_1,\gamma_2,...,\gamma_{2N-1})$
the space of operators is of dimension 
\begin{eqnarray}
{\cal N}_{op} = 2^{2N-1}
\label{nop}
\end{eqnarray}
and can be decomposed into the even and the odd subspaces of equal dimensions
\begin{eqnarray}
{\cal N}^{odd}_{even}={\cal N}^{odd}_{op} && = \frac{{\cal N}_{op}}{2}= 2^{2N-2}
\label{noddop}
\end{eqnarray}
The standard inner product between two operators $X$ and $Y$ reads in terms of the normalized trace ${\rm tr}$
\begin{eqnarray}
( X,Y ) \equiv \frac{{\rm Tr} (X^{\dagger} Y)}{{\rm Tr}  (\mathbb{1} )}  =\frac{{\rm Tr} (X^{\dagger} Y)}{ 2^{2N-1} }  \equiv  {\rm tr} (X^{\dagger} Y)
\label{inner}
\end{eqnarray}

It is convenient to associate to any odd number $(2k-1)$ with $k=1,2,..N$ of Majorana operators labelled by $1 \leq j_1<j_2<..<j_{2k-1} \leq 2N-1$
the operator \cite{goldstein} 
\begin{eqnarray}
{\Upsilon}^{(2k-1)}_{j_1,j_2,..,j_{2k-1}} \equiv   i^{k-1} \gamma_{j_1} \gamma_{j_2}\gamma_{j_3} \gamma_{j_4}  ... \gamma_{j_{2k-1}} 
\label{upsilonj}
\end{eqnarray}
For $k=1$ one recovers the individual Majorana operators
\begin{eqnarray}
{\Upsilon}^{(1)}_{j_1}  && =  \gamma_{j_1} 
\label{ups1}
\end{eqnarray}
while for $k=2$ and $k=3$, they read respectively
\begin{eqnarray}
{\Upsilon}^{(3)}_{j_1,j_2,j_3}  && =   i \gamma_{j_1} \gamma_{j_2} \gamma_{j_3}
\nonumber \\
{\Upsilon}^{(5)}_{j_1,j_2,j_3,j_4,j_5} &&= - \gamma_{j_1} \gamma_{j_2} \gamma_{j_3} \gamma_{j_4}\gamma_{j_5}
\label{ups35}
\end{eqnarray}
Finally for $k=N$ the only possibility $j_q=q$ corresponds to the operator $\Upsilon^{tot}$ already introduced in Eq. \ref{upstot}.

The operators of Eq. \ref{upsilonj} are hermitian
\begin{eqnarray}
( {\Upsilon}^{(2k-1)}_{j_1,j_2,..,j_{2k-1}}  )^{\dagger} = \Upsilon^{(2k-1)}_{j_1,j_2,..,j_{2k-1}} 
\label{parityjh}
\end{eqnarray}
square to the Identity
\begin{eqnarray}
({\Upsilon}^{(2k-1)}_{j_1,j_2,..,j_{2k-1}}  )^2    = \mathbb{1}
\label{parityjsquare}
\end{eqnarray}
and form the standard orthonormal basis of the odd subspace \cite{goldstein}.

\subsection{ Reminder on the Goldstein-Chamon matrix within the odd subspace \cite{goldstein}  }

Let us relabel the orthonormal basis ${\Upsilon}_{\mu}$
with the single index $\mu=1,..,2^{2N-2}$.
The dynamics of ${\Upsilon}_{\mu} $ is given by the Heisenberg equation that can be projected on this basis
\begin{eqnarray}
\frac{ d {\Upsilon}_{\mu}}{dt} = i [H, {\Upsilon}_{\mu}] =i  \sum_{\nu} \Upsilon_{\nu} {\cal H}_{\nu,\mu} 
\label{dynups}
\end{eqnarray}
where the Goldstein-Chamon matrix \cite{goldstein}  is defined in terms of the inner product of Eq. \ref{inner}
\begin{eqnarray}
{\cal H}_{\nu,\mu} \equiv   ( \Upsilon_{\nu},[H,   {\Upsilon}_{\mu}] ) =   {\rm tr} ( \Upsilon_{\nu} H   {\Upsilon}_{\mu}
- \Upsilon_{\nu}{\Upsilon}_{\mu} H  )
\label{gcmatrix}
\end{eqnarray}
This matrix is antisymmetric (as a consequence of the cyclic invariance of the trace)
\begin{eqnarray}
{\cal H}_{\nu,\mu}  =  {\rm tr} ( {\Upsilon}_{\mu} \Upsilon_{\nu} H   - {\Upsilon}_{\mu} H  \Upsilon_{\nu} )
= -  ( \Upsilon_{\mu},[H,   {\Upsilon}_{\nu}] ) 
\label{hmatrix}
\end{eqnarray}
and can be also rewritten as (again using the cyclic invariance of the trace)
\begin{eqnarray}
{\cal H}_{\nu,\mu}  =  {\rm tr} (  H   {\Upsilon}_{\mu}\Upsilon_{\nu}
-  H \Upsilon_{\nu}{\Upsilon}_{\mu}   ) = ( H ,[ \Upsilon_{\mu},   {\Upsilon}_{\nu}] ) 
\label{heisenberg}
\end{eqnarray}
so that it vanishes when the commutator $[ \Upsilon_{\mu},   {\Upsilon}_{\nu}]  $ is zero.
The Goldstein-Chamon matrix  ${\cal H}_{\nu,\mu} $ can of course be written similarly in the even subspace \cite{goldstein},
but will not be discussed here.

\subsection{ Flavor of odd operators}

Now we wish to adapt the framework described above 
 to the presence of the Time-Reversal Symmetry $T$ (Eq \ref{time}).
The Hamiltonian involves
the $N$ Majorana operators $a_i=\gamma_{2i-1}$ with $i=1,..,N$
and the $(N-1)$  Majorana operators $b_i=\gamma_{2i}$ with $i=1,..,N-1$ that behave differently with respect to the Time-Reversal Symmetry $T$ (Eq \ref{time}).

If  the odd operator of Eq. \ref{upsilonj} contains $n_a$ operators $a_j=\gamma_{2j-1}$ and $n_b$ operators $b_j=\gamma_{2j}$ 
with $2k-1=n_a+n_b$, the time reversal action becomes (Eq \ref{time})
\begin{eqnarray}
T {\Upsilon}^{(2k-1)}_{j_1,j_2,..,j_{2k-1}} T^{-1} =    (-1)^{k+1+n_b} {\Upsilon}^{(2k-1)}_{j_1,j_2,..,j_{2k-1}}
=   (-1)^{\frac{n_a-n_b-1}{2}} {\Upsilon}^{(2k-1)}_{j_1,j_2,..,j_{2k-1}}
\label{tupsilonj}
\end{eqnarray}
It is then useful to separate the odd operators $\Upsilon_{\mu}$
into operators $A_{\alpha}$ of flavor A (sector $T=+1$ ) and operators $B_{\beta}$ flavor B (sector $T=-1$ ).

\subsubsection{ Odd Operators of flavor $A$ (sector $T=+1$ )}

The odd Operators of the sector $T=+1$ corresponds to $n_a-n_b=4 m+1$ while $n_a+n_b=2k-1$, so the possible cases
\begin{eqnarray}
0 \leq n_a && =k+2m \leq N
\nonumber \\
0 \leq n_b && =k-2m-1 \leq N-1
\label{flavora}
\end{eqnarray}
are labelled by the two integers $(k,m)$.
The possible values of $k$ are
\begin{eqnarray}
k=1,2,..., N
\label{oddnanb}
\end{eqnarray}
For each value of $k$, the possible values of $m$ are given by
\begin{eqnarray}
{\rm max} (-k, -(N-k) ) \leq 2m \leq {\rm min} (k-1,N-k)
\label{oddnanbm}
\end{eqnarray}

For $k=1$ corresponding to a single operator (Eq \ref{ups1})
 the only possibility is $m=0$ i.e. $(n_a=1,n_b=0)$ that corresponds as it should to the $N$ Majorana operators $a_i$
\begin{eqnarray}
A_i= a_i \ \ {\rm for \ \ }  i=1,2,..,N
\label{k1a}
\end{eqnarray}

For $k=N$ corresponding to all the $(2N-1)$ operators (Eq. \ref{upstot}) i.e $n_a=N$ and $n_b=N-1$,
one recovers the operator ${\Upsilon}^{tot}  $ of Eq. \ref{upstot}.

\subsubsection{ Odd Operators of flavor $B$ (sector $T=-1$ )  }

The odd Operators of the sector $T=-1$ correspond to $n_a-n_b=4 m-1$ while $n_a+n_b=2k-1$, so the possible cases
\begin{eqnarray}
 0 \leq n_a && =k+2m-1   \leq N
\nonumber \\
0 \leq n_b && =k-2m   \leq N-1
\label{flavorb}
\end{eqnarray}
are labelled by the two integers $(k,m)$.

The possible values of $k$ are
\begin{eqnarray}
k=1,2,..., N-1
\label{oddnanbkb}
\end{eqnarray}
For each value of $k$, the possible values of $m$ are given by
\begin{eqnarray}
{\rm max} (-(k-1), -(N-k-1) ) \leq 2m \leq {\rm min} (k,N+1-k)
\label{oddnanbkm}
\end{eqnarray}

For $k=1$ corresponding to a single operator (Eq \ref{ups1})
 the only possibility is $m=0$ i.e. $(n_a=0,n_b=1)$ that corresponds as it should to the $N-1$ Majorana operators $b_i$
\begin{eqnarray}
B_i= b_i \ \ {\rm for \ \ }  i=1,2,..,N-1
\label{k1b}
\end{eqnarray}

\subsubsection{ Dimensions of the subspaces of flavors $A$ and $B$  }

For a given number $(2k-1)$ of operators, the total number of operators of any flavor
 is given by the binomial number of choices of $(2k-1)$ operators among $(2N-1)$
\begin{eqnarray}
  {\cal N}_{op}^{odd A (2k-1) } + {\cal N}_{op}^{odd B (2k-1) } = {\cal N}^{odd (2k-1) }_{op} = \binom{2N-1}{2k-1} 
\label{calnsum}
\end{eqnarray}
while the difference between the two flavors $A$ and $B$ can be evaluated to be
\begin{eqnarray}
  {\cal N}_{op}^{odd A (2k-1) } - {\cal N}_{op}^{odd B (2k-1) } = {\cal D}^{odd (2k-1) }_{op}=    \binom{N-1}{k-1} 
\label{calndiff}
\end{eqnarray}
so that one obtains respectively the dimensions of the subspaces of flavors $A$ and $B$ for a given number $(2k-1)$ of operators
\begin{eqnarray}
  {\cal N}_{op}^{odd A (2k-1) } && = \frac{  {\cal N}^{odd (2k-1) }_{op} +{\cal D}^{odd (2k-1) }_{op} }{2} 
 = \frac{  \binom{2N-1}{2k-1} + \binom{N-1}{k-1} }{2}
\nonumber \\
 {\cal N}_{op}^{odd B (2k-1) } && = \frac{  {\cal N}^{odd (2k-1) }_{op} - {\cal D}^{odd (2k-1) }_{op} }{2} 
 = \frac{  \binom{2N-1}{2k-1} - \binom{N-1}{k-1} }{2}
\label{calnabk}
\end{eqnarray}

As a consequence, the total numbers of operators of flavors $A$ and $B$ are given by
\begin{eqnarray}
 {\cal N}_{op}^{odd A  } && = \sum_{k=1}^N {\cal N}_{op}^{odd A (2k-1) } 
 = \frac{  2^{2N-2}+ 2^{N-1} }{2} = 2^{2N-3}+ 2^{N-2}
\nonumber \\
  {\cal N}_{op}^{odd B  } && = \sum_{k=1}^N {\cal N}_{op}^{odd B (2k-1) } 
 = \frac{  2^{2N-2}- 2^{N-1} }{2} = 2^{2N-3}- 2^{N-2}
\label{calnanb}
\end{eqnarray}
In particular, the difference between the two dimensions reads
\begin{eqnarray}
{\cal D}^{odd }_{op} = {\cal N}_{op}^{odd A  }-  {\cal N}_{op}^{odd B  }&& =    2^{N-1}
\label{diffab}
\end{eqnarray}

\subsubsection{ Adaptation of the Goldstein-Chamon matrix  }

In the presence of the Time-Reversal-Symmetry $T$, it is easy to see
that the Goldstein-Chamon matrix ${\cal H}_{\mu\nu}$ within the odd subspace (Eq. \ref{gcmatrix})
vanishes between two odd operators of the same flavor.
It is then convenient to reshape
the Goldstein-Chamon matrix ${\cal H}_{\mu\nu}$ 
into the following real rectangular matrix
 \begin{eqnarray}
M_{\beta \alpha} \equiv \frac{1}{2i} ( B_{\beta}, [H, A_{\alpha}]) =- \frac{1}{2i} (A_{\alpha} , [H, B_{\beta}]) = \frac{1}{2i} ( H, [ A_{\alpha},B_{\beta}]) 
\label{mbetaalpha}
\end{eqnarray}
of size
\begin{eqnarray}
  {\cal N}_{op}^{odd B  }  \times  {\cal N}_{op}^{odd A  } && = ( 2^{2N-3}- 2^{N-2}) \times ( 2^{2N-3}+ 2^{N-2} )
\label{totoddcardab}
\end{eqnarray}

In terms of this matrix $M_{\beta\alpha}$ (Eq. \ref{mbetaalpha}), the dynamics of the odd operators of flavor $B$ reads
\begin{eqnarray}
\frac{ d B_{\beta}}{dt} = i [H,B_{\beta} ] = 2  \sum_{\alpha}  M_{\beta \alpha} A_{\alpha}
\label{dynupsb}
\end{eqnarray}
while the dynamics of the odd operators of flavor $A$ reads
\begin{eqnarray}
\frac{ d A_{\alpha}}{dt} = i [H, A_{\alpha}] = -2 \sum_{\beta} B_{\beta}   M_{\beta \alpha}
\label{dynupsa}
\end{eqnarray}

To obtain closed dynamical equations within the sector of flavor $B$, it is thus convenient to write the second time derivatives to obtain
\begin{eqnarray}
\frac{ d^2 B_{\beta}}{dt^2} = -4  \sum_{\beta'} N_{\beta \beta'} B_{\beta'}  
\label{dynupsb2}
\end{eqnarray}
in terms of the symmetric real square matrix of size ${\cal N}_{op}^{odd B  }  \times {\cal N}_{op}^{odd B  }    $
\begin{eqnarray}
N_{\beta \beta'} = (M M^t )_{\beta \beta'} =\sum_{\alpha}   M_{\beta \alpha} M_{\beta' \alpha}
\label{nmatrix}
\end{eqnarray}
To see how this general formalism works in practice, 
it is now useful to study small systems where the Hamiltonian depends only on $2N-1=3$ and $2N-1=5$ Majorana fermions.

\section{ Example with $2N-1=3$ Majorana fermions  }

\label{sec_three}

The Hamiltonian respecting the Parity $P$, the Time-Reversal-Symmetry $T$
depends only the three Majorana operators $(a_1=\gamma_1,b_1=\gamma_2,a_2=\gamma_3)$,
so that it can only involves two couplings $K_1$ and $K_2$
\begin{eqnarray}
H= i K_1 a_1 b_1 + i K_2 b_1 a_2 = i b_1 (- K_1 a_1 +K_2 a_2)
\label{h3}
\end{eqnarray}
The translation in the spin language is given in Eq. \ref{h3spin} of the Appendix.
Even if this case is too small to contain four-Majorana-fermions interactions, 
it is nevertheless useful to mention how the various notions described above apply in such a simple case.

\subsection { Diagonalization in terms of pseudo-Majorana fermions }

Here the diagonalization of Eq \ref{h3} in terms of pseudo-Majorana fermions is of course completely obvious.
One just needs to replace the two Majorana operators $(a_1,a_2)$ of flavor $A$ by the new Majorana operators $(\tilde a_1,\tilde a_2)$
obtained by the rotation
\begin{eqnarray}
\tilde a_1 && =  \cos\theta a_1 + \sin \theta a_2
\nonumber \\
\tilde a_2 &&= - \sin \theta a_1 + \cos \theta a_2 
\label{13atilde}
\end{eqnarray}
of angle $\theta$ defined by
\begin{eqnarray}
 \cos \theta = \frac{K_2}{\sqrt{ K_1^2+K_2^2} }
\nonumber \\
 \sin \theta = \frac{K_1}{\sqrt{ K_1^2+K_2^2} }
\label{theta}
\end{eqnarray}
so that the Hamiltonian of Eq. \ref{h3} reduces to
\begin{eqnarray}
H= i \sqrt{ K_1^2+K_2^2}  b_1  \tilde a_2 
\label{H3tilde}
\end{eqnarray}
and does not involve $ \tilde a_1$.

\subsection { Even Zero Modes }

The $ 2^{N-1}=2$ even zero-modes of Eq. \ref{evenpower} are given by
\begin{eqnarray}
Z^{even}_{p=0} && =   \mathbb{1}
\nonumber \\
Z^{even}_{p=1} && =   H =  i b_1 (-K_1 a_1 + K_2 a_2)
\label{evenzeromode}
\end{eqnarray}
while the next power $p=2$ of the Hamiltonian gives a constant 
\begin{eqnarray}
H^2=K_1^2+K_2^2
\label{HKparity}
\end{eqnarray}
and thus is not linearly independent of $Z^{even}_{p=0}  =   \mathbb{1} $.

The normalized even zero mode that appear in the Hamiltonian is the pseudo-spin
\begin{eqnarray}
\tau^z_1 =i   b_1  \tilde a_2  = \frac{ H}{ \sqrt{ K_1^2+K_2^2} }
\label{tauz1h3}
\end{eqnarray}
while the other pseudo-spin absent from the Hamiltonian (Eq \ref{deftaun} is
\begin{eqnarray}
\tau^z_2 =i   b_2  \tilde a_1  
\label{tauz2h3}
\end{eqnarray}

\subsection { Odd Zero Modes }

The $ 2^{N-1}=2$ odd zero-modes of Eq. \ref{zerooddp} are given by
\begin{eqnarray}
Z^{odd}_{p=0} && =  {\Upsilon}^{tot} =  i a_1 b_1 a_2 = i \tilde a_1 b_1 \tilde a_2 =  \tilde a_1 \tau^z_1
\nonumber \\
Z^{odd}_{p=1} && = H {\Upsilon}^{tot} = K_2 a_1 + K_1 a_2 =  \sqrt{ K_1^2+K_2^2} \tilde a_1
\label{zerooddp01}
\end{eqnarray}

\subsection { Dynamics of Majorana operators }

Here, the number ${\cal N}_{op}^{odd}=2^{2N-2}=4$ of odd operators decomposes into ${\cal N}_{op}^{oddA}=2^{2N-3}+2^{N-2}=2+1=3 $ operators of flavor $A$
\begin{eqnarray}
A_1&& = a_1
\nonumber \\
A_2&& = a_2
\nonumber \\
A_3 && = {\Upsilon}^{tot} =  i a_1 b_1 a_2
\label{a123}
\end{eqnarray}
and ${\cal N}_{op}^{oddB}=2^{2N-3}-2^{N-2}=2-1=1 $ operator of flavor $B$
\begin{eqnarray}
B_1&& = b_1
\label{b1}
\end{eqnarray}
whose dynamics
\begin{eqnarray}
\frac{ d b_1}{dt} = i [H,b_1 ] = 2 (-K_1 a_1 + K_2 a_2) 
\label{dynupsb1}
\end{eqnarray}
yields that the real rectangular matrix $M_{\beta \alpha}$ (Eq. \ref{dynupsb}) of dimension $1 \times 3$ is simply
\begin{eqnarray}
 M = (-K_1,K_2,0)
\label{m3}
\end{eqnarray}

\subsection { Links with various Real Space Renormalization procedures for spin models }

The above analysis actually provides a new interesting Majorana-interpretation of the self-dual
Pacheco-Fernandez block-spin renormalization for the ground-state of the pure or random quantum Ising model  \cite{pacheco,igloiSD,nishiRandom,us_pacheco,us_renyi}, where the intra-block Hamiltonian for two spins
corresponds to Eq. \ref{h3} in the spin language of Eq. \ref{h3spin} : the renormalization procedure consists in projecting
the pseudo-spin $\tau^z_1 $ of Eq. \ref{tauz1h3} into the value that minimizes the intra-block Hamiltonian
\begin{eqnarray}
\tau^z_1 = -1
\label{tauz1h3gs}
\end{eqnarray}
while the other pseudo spin $\tau^z_2 $ absent from the intra-block Hamiltonian (Eq \ref{deftaun})
that labels the two degenerate ground-states of the intra-block Hamiltonian 
is kept as the renormalized spin for the block. The inter-block Hamiltonian is then taken into account to compute
the renormalized interactions between these renormalized spins (see \cite{pacheco,igloiSD,nishiRandom,us_pacheco,us_renyi}
for more details).
In the random case, this block real-space renormalization for the ground-state can be promoted to a block real-space renormalization for all the eigenstates
by projecting the pseudo-spin of Eq. \ref{tauz1h3} into its two possible values 
\begin{eqnarray}
\tau^z_1 = \pm 1
\label{tauz1h3x}
\end{eqnarray}
in each block, as discussed in detail in \cite{c_emergent}, in relation with the RSRG-X procedures 
introduced to construct the whole set of excited eigenstates in Many-Body-Localized phase \cite{rsrgx,rsrgx_moore,vasseur_rsrgx,yang_rsrgx,rsrgx_bifurcation,c_rsrgxMaj},
or with the related RSRG-t procedure to describe the corresponding effective dynamics \cite{vosk_dyn1,vosk_dyn2,c_rsrgt}.
The analogous block renormalization procedure for the Floquet dynamics in Many-Body-Localized phase can be found in \cite{c_floquet}.
Besides the application to all blocks in parallel just described, this renormalization procedure can be instead applied sequentially
in order to generalize it to other geometries like stars and watermelons \cite{us_watermelon} or the Cayley tree \cite{c_mblcayley}.

\subsection { Real-space renormalization based on blocks of three Majorana fermions }

Besides the nice re-interpretation of the previously known renormalization schemes for quantum spin chains that we have just described,
the Majorana formulation suggests new real-space renormalization procedures based on blocks of an odd number of Majorana fermions.
For instance for the quantum Ising model discussed above that translates into the following Kitaev chain (see Appendix)
\begin{eqnarray}
H && = i \sum_n K_n \gamma_n \gamma_{n+1 } 
\nonumber \\
&& = i K_1 a_1 b_1 + i K_2 b_1 a_2 + i K_3 a_2 b_2 + i K_4 b_2 a_3 + i K_5 a_3 b_3 +
+ i K_6  b_3 a_4 + i K_7  a_4 b_4 + i K_8 b_4 a_5+...
\label{hkitaev}
\end{eqnarray}
it would seem more natural to divide the chain into blocks of three Majorana fermions.
The first block concerning $(a_1,b_1,a_2)$ would have for internal Hamiltonian Eq .\ref{h3}
\begin{eqnarray}
H^{intra}_{block1}=  i K_1 a_1 b_1 + i K_2 b_1 a_2 
\label{hkitaev1}
\end{eqnarray}
the second block concerning $(b_2, a_3, b_3)$ would have for internal Hamiltonian
\begin{eqnarray}
H^{intra}_{block2}=   i K_4 b_2 a_3 + i K_5 a_3 b_3 
\label{hkitaev2}
\end{eqnarray}
while the inter-Hamiltonian between these two blocks would be
\begin{eqnarray}
H^{inter}_{block12}=  i K_3 a_2 b_2 
\label{hkitaev12}
\end{eqnarray}
In the renormalization procedure, the first block is replaced by the renormalized Majorana fermion $\tilde a_1$ of Eq. \ref{13atilde}
\begin{eqnarray}
a^R_{block1}=   \frac{K_2 a_1 + K_1 a_2 }{\sqrt{ K_1^2+K_2^2} }
\label{arblock1}
\end{eqnarray}
the second block is replaced by the renormalized Majorana fermion of flavor $b$
\begin{eqnarray}
b^R_{block2}=   \frac{K_5 b_2 + K_4 b_3 }{\sqrt{ K_4^2+K_5^2} }
\label{brblock2}
\end{eqnarray}
 and so on, while the inter-block Hamiltonian of Eq. \ref{hkitaev12}
produces the following renormalized coupling between the renormalized Majorana fermions of the two first blocks
\begin{eqnarray}
K^{R}_{block1,block2}= \frac{ K_1  }{\sqrt{ K_1^2+K_2^2} }  K_3 \frac{K_5  }{\sqrt{ K_4^2+K_5^2} }
\label{kr12}
\end{eqnarray}
Similarly, the renormalized coupling between the second and the third blocks reads
\begin{eqnarray}
K^{R}_{block2,block3}= \frac{ K_4  }{\sqrt{ K_4^2+K_5^2} }  K_6 \frac{K_8  }{\sqrt{ K_7^2+K_8^2} }
\label{kr23}
\end{eqnarray}
These renormalization rules are thus very similar to the usual Pacheco-Fernandez rules \cite{pacheco,igloiSD,nishiRandom,us_pacheco,us_renyi}
but are more symmetric because the two Majorana flavors $a$ and $b$ are treated on the same footing.
As in \cite{us_pacheco}, these rules correspond to an Infinite-Disorder-Fixed-Point of activated exponent $\psi=1/2$ and correlation exponent $\nu=1$,
in agreement with the exact solution of Daniel Fisher \cite{fisher} 
obtained by the Strong Disorder RG approach (see the review \cite{strong_review}).

In conclusion, from a real-space renormalization perspective,
it can be advantageous to replace the notion of blocks containing an integer number of spins (corresponding to an even number of Majorana fermions)
by the notion of blocks containing an odd number of Majorana fermions (corresponding to an half-integer number of spins !)
so that the renormalized Majorana fermion for the block corresponds to the quasi Majorana fermion absent from the intra-block Hamiltonian.

\section{ Example with $2N-1=5$ Majorana fermions  }

\label{sec_five}

The Hamiltonian respecting the Parity $P$ and the Time-Reversal-Symmetry $T$  
depends only on three Majorana fermions of flavor $A$ (  $a_1=\gamma_1$, $a_2=\gamma_3$, $a_3=\gamma_5$ )
and two Majorana fermions of flavor $B$ ( $b_1=\gamma_2$ , $b_2=\gamma_4$ )
so that it can involve $9$ couplings 
that can be labelled by three vectors $(\vec I, \vec J, \vec G)$ of three components each
\begin{eqnarray}
H  && = i b_1 \sum_{j=1}^3 I_j a_j +  i b_2 \sum_{j=1}^3 J_j a_j + b_1b_2 ( G_1 a_2 a_3- G_2 a_1 a_3 + G_3 a_1 a_2)
\label{h5}
\end{eqnarray}
The translation in the spin language is given in Eq. \ref{h5spin} of the Appendix.

To obtain more explicit final results, it will be sometimes convenient to focus on the following special case with only 5 non-vanishing couplings
\begin{eqnarray}
0= I_1=I_3 =J_2=G_2
\label{special5}
\end{eqnarray}
(see Eq. \ref{h5spinspecial} for the translation in the spin language).

\subsection{ Even zero modes given by the first powers of the Hamiltonian}

The ${\cal N} = 2^{N-1}=4$ first powers of the Hamiltonian  labelled by $p=0,1,2,3$ ( Eq. \ref{evenpower}) can be decomposed into
\begin{eqnarray}
Z^{even}_{p} && =    H^p   = t_p +  {\cal H}_p
\label{evenzeromode5}
\end{eqnarray}
where one separates the constant contribution given by the normalized trace
\begin{eqnarray}
t_p &&  \equiv tr(H^p) = \frac { Tr(H^p)}{Tr( \mathbb{1} )} 
\label{tpnormalizedtrace}
\end{eqnarray}
while the trace-less part that has the same form as the initial Hamiltonian of Eq. \ref{h5}
with its own couplings
\begin{eqnarray}
 {\cal H}_p && \equiv  i b_1 \sum_{j=1}^3 I^{(p)}_j a_j +  i b_2 \sum_{j=1}^3 J_j^{(p)} a_j + b_1b_2 ( G_1^{(p)} a_2 a_3- G_2^{(p)} a_1 a_3 + G_3^{(p)} a_1 a_2)
\label{calhp}
\end{eqnarray}
For $p=0,1$ these notations mean $t_0=1$ , $ {\cal H}_0=0$ and  $t_1=0$, $ {\cal H}_1=H$ respectively
\begin{eqnarray}
Z^{even}_{p=0} && =  \mathbb{1}
\nonumber \\
Z^{even}_{p=1} && = H =  i b_1 \sum_{j=1}^3 I_j a_j +  i b_2 \sum_{j=1}^3 J_j a_j + b_1b_2 ( G_1 a_2 a_3- G_2 a_1 a_3 + G_3 a_1 a_2)
\label{evenzeromode01maj}
\end{eqnarray}

For $p=2$ and $p=3$,  the computation yields respectively
\begin{eqnarray}
t_2&& = \sum_{j=1}^3 (I_j^2+J_j^2+G_j^2)  =     \Vert \vec I  \Vert^2+ \Vert \vec J  \Vert^2 + \Vert \vec G  \Vert^2
\nonumber \\
\vec I^{(2)} && = 2 \vec J \times \vec G
\nonumber \\
\vec J^{(2)} && = 2 \vec G \times \vec I
\nonumber \\
\vec G^{(2)} && = 2 \vec I \times \vec J
\label{couph2}
\end{eqnarray}
and
\begin{eqnarray}
 t_3 && =   \vec I .  \vec I^{(2)} +  \vec J .  \vec J^{(2)} +  \vec G .  \vec G^{(2)} = 6 det(\vec I, \vec J,\vec G)
\nonumber \\
\vec I^{(3)} && = t_2 \vec I + \vec J \times \vec G^{(2)} -  \vec G \times \vec J^{(2)} 
= 3  t_2 \vec I  - 2 \Vert \vec I  \Vert^2 \vec I -2 (\vec I. \vec J) \vec J - 2  (\vec I. \vec G) \vec G
\nonumber \\
\vec J^{(3)} && =  t_2 \vec J  +  \vec G \times \vec I^{(2)} - \vec I \times \vec G^{(2)}  
=3  t_2 \vec J   -2 \Vert \vec J \Vert^2 \vec J -2 (\vec J. \vec I) \vec I - 2  (\vec J. \vec G) \vec G
\nonumber \\
\vec G^{(3)} && =  t_2 \vec G+\vec I \times \vec J^{(2)} - \vec J \times \vec I^{(2)} 
=3  t_2 \vec G  -2  \Vert \vec G  \Vert^2 \vec G - 2  (\vec G. \vec I) \vec I-2 (\vec G. \vec J) \vec J 
\label{couph3}
\end{eqnarray}

For the next power $p=4$, the evaluation of  $H^4=(H^2)^2$ yields
\begin{eqnarray}
 t_4&& = t_2^2 +
 \Vert \vec I^{(2)}  \Vert^2+ \Vert \vec J^{(2)}  \Vert^2 + \Vert \vec G^{(2)}  \Vert^2
\nonumber \\
&& =t_2^2+ 4 (  \Vert \vec I  \Vert^2 \Vert \vec J  \Vert^2 - (\vec I. \vec J)^2 
+  \Vert \vec I  \Vert^2 \Vert \vec G  \Vert^2 - (\vec I. \vec G)^2 
+  \Vert \vec J  \Vert^2 \Vert \vec G  \Vert^2 - (\vec J. \vec G)^2   )
\nonumber \\
\vec I^{(4)} && = 2 t_2\vec I^{(2)}   +   2 \vec J^{(2)} \times \vec G^{(2)}
=2 t_2\vec I^{(2)} +   8 det(\vec I, \vec J,\vec G) \vec I 
\nonumber \\
\vec J^{(4)} && = 2t_2 \vec J^{(2)}  + 2 \vec G^{(2)} \times \vec I^{(2)}
=2 t_2 \vec J^{(2)} +  8 det(\vec I, \vec J,\vec G) \vec J
\nonumber \\
\vec G^{(4)} && = 2t_2 \vec G^{(2)}   + 2 \vec I^{(2)} \times \vec J^{(2)}
=2 t_2 \vec G^{(2)} +  8 det(\vec I, \vec J,\vec G) \vec G
\label{couph4}
\end{eqnarray}
so that it can be rewritten as the following linear combination of the three lower powers $(  \mathbb{1},H,H^2)$ as
\begin{eqnarray}
 H^4 =2 t_2 H^2+ \frac{4}{3} t_3   H + (t_4-2 t_2^2)
\label{cayleyh4}
\end{eqnarray}
in agreement with the Cayley-Hamilton theorem recalled in Eq \ref{cayleyhamilton}.

\subsection{ Normalized even zero modes }

The diagonalization in terms of two pseudo-spins of Eq. \ref{hmbl}
reads with $\omega_0=0$ and the relabelling $\omega_{12} \to \omega_3$
\begin{eqnarray}
 H && =  \omega_1 \tau_1^z+   \omega_2 \tau_2^z + \omega_3 \tau_1^z \tau_2^z
\nonumber \\
&&=  \omega_1 (i \tilde b_1 \tilde a_2) +\omega_2 (i \tilde b_2 \tilde a_3)   + \omega_3  ( \tilde b_1  \tilde b_2 \tilde a_2 \tilde a_3)  
\label{h3pseudo}
\end{eqnarray}
This form would correspond for Eq. \ref{h5} to the pseudo-couplings
\begin{eqnarray}
\tilde I_2=\omega_1 
\nonumber \\
\tilde J_3=\omega_2
\nonumber \\
\tilde G_1=\omega_3
\label{hpseudoc}
\end{eqnarray}
while the six other pseudo-couplings vanish.

As a consequence, we may compute the traces of the first powers of the Hamiltonian with the general formula derived above to obtain
the symmetric polynomials of $(\omega_1^2,\omega_2^2,\omega_3^2)$ in terms of the initial couplings of Eq. \ref{h5}
\begin{eqnarray}
E_1 && \equiv     \omega_1^2+   \omega_2^2 + \omega_3^2  = t_2= \Vert \vec I  \Vert^2+ \Vert \vec J  \Vert^2 + \Vert \vec G  \Vert^2
\nonumber \\
 E_2 && \equiv  \omega_1^2   \omega_2^2 + \omega_1^2 \omega_3^2+   \omega_2^2  \omega_3^2
= \frac{ t_4-t_2^2 }{4} = \Vert \vec I  \Vert^2 \Vert \vec J  \Vert^2 - (\vec I. \vec J)^2 
+  \Vert \vec I  \Vert^2 \Vert \vec G  \Vert^2 - (\vec I. \vec G)^2 
+  \Vert \vec J  \Vert^2 \Vert \vec G  \Vert^2 - (\vec J. \vec G)^2
\nonumber \\
 E_3 && \equiv \omega_1^2 \omega_2^2 \omega_3^2 =\left( \frac{t_3}{6} \right)^2 = (det(\vec I, \vec J,\vec G) )^2
\label{eom}
\end{eqnarray}
The squares $(\omega^2_1, \omega_2^2 , \omega_3^2 )$
of the three pseudo-couplings of Eq. \ref{h3pseudo}
may be thus obtained as the three roots of the following cubic equation in the variable $x=\omega^2$
\begin{eqnarray}
0&& = (x-    \omega_1^2) (x-   \omega_2^2) (x- \omega_3^2) 
 \nonumber \\
&&  = x^3 - E_1 x^2  + x E_2 -  E_3
\label{cubic}
\end{eqnarray}

Once these pseudo-couplings have been computed, the pseudo spins $\tau_1^z $ and $\tau^z_2$ may be obtained from the identification of
the three first trace-less powers of the Hamiltonian
\begin{eqnarray}
{\cal H}_1\equiv H && =  \omega_1 \tau_1^z+   \omega_2 \tau_2^z + \omega_3 \tau_1^z \tau_2^z
\nonumber \\
{\cal H}_2\equiv  H^2-t_2 && = ( 2 \omega_2 \omega_3) \tau_1^z +  ( 2 \omega_1 \omega_3) \tau_2^z +  ( 2 \omega_1 \omega_2) \tau_1^z\tau_2^z
\nonumber \\
{\cal H}_3\equiv H^3-t_3&& =  \omega_1 (3 t_2 - \omega_1^2) \tau_1^z+   \omega_2 (3 t_2 - \omega_2^2)  \tau_2^z 
+ \omega_3 (3 t_2 - \omega_3^2)  \tau_1^z \tau_2^z
\label{hpseudo}
\end{eqnarray}
This system of three equations can be written in a simpler form by transforming the the second and third equations into
\begin{eqnarray}
 3 \frac{{\cal H}_2 }{  t_3 } && =\frac{1}{\omega_1} \tau_1^z +\frac{1}{\omega_2}  \tau_2^z +\frac{1}{\omega_3}   \tau_1^z\tau_2^z
\nonumber \\
 \frac{(3t_2 {\cal H}_1 - {\cal H}_3) }{2} && =  \omega_1^3 \tau_1^z+   \omega_2^3 \tau_2^z + \omega_3^3 \tau_1^z \tau_2^z
\label{hpseudobis}
\end{eqnarray}
The solution of this linear system yields the two pseudo-spins $\tau_1^z $ and $\tau^z_2$ in terms of the first three traceless powers of the Hamiltonian
\begin{eqnarray}
\tau_k^z =\frac{1}{2 D_k} \left( g_{k1}   {\cal H}_1 +g_{k2} {\cal H}_2 +g_{k3}  {\cal H}_3 \right)
\label{pseudospin}
\end{eqnarray}
with the notations
\begin{eqnarray}
g_{k1} && = \omega_k \left( 2 \omega_k^2+ t_2 \right)
\nonumber \\
g_{k2} && = \frac{t_3}{ 6 \omega_k }
\nonumber \\
g_{k3} && =-  \omega_k 
\nonumber \\
D_k && = \prod_{j \ne k} (\omega_k^2-\omega_j^2) = \omega_k^4- \omega_k^2 (t_2-  \omega_k^2) + \frac{E_3}{ \omega_k^2}
\label{coefspseudo}
\end{eqnarray}

To obtain more explicit final results than the application to Eq. \ref{cubic} of the general formula  known for cubic equations, 
it is now useful to focus on the special case of Eq. \ref{special5}.
Then the symmetric polynomials of Eq. \ref{eom} reduce to
\begin{eqnarray}
E_1 && = I_2^2+ (J_1^2+J_3^2) + (G_1^2+G_3^2)
\nonumber \\
 E_2 && = I_2^2 (  (J_1^2+J_3^2) + (G_1^2+G_3^2) ) +  ( J_1 G_3-G_1 G_3)^2
\nonumber \\
 E_3 &&  = I_2^2  ( J_1 G_3-J_3 G_1)^2
\label{eomspecial}
\end{eqnarray}
So the cubic equation of Eq. \ref{cubic} has one trivial root
\begin{eqnarray}
\omega_1^2 && = x_1 = I_2^2
\label{cubic1}
\end{eqnarray}
The two other roots have for sum and products
\begin{eqnarray}
\omega_2^2+ \omega_3^2 && =(J_1^2+J_3^2) + (G_1^2+G_3^2) 
\nonumber \\
\omega_2^2 \omega_3^2  && = ( J_1 G_3-J_3 G_1)^2
\label{cubic23}
\end{eqnarray}
and one obtains
\begin{eqnarray}
\omega_2^2 && = x_2 = \frac{(J_1^2+J_3^2) + (G_1^2+G_3^2) + \sqrt{\Delta}}{2}
\nonumber \\
\omega_3^2 && = x_3 = \frac{(J_1^2+J_3^2) + (G_1^2+G_3^2) - \sqrt{\Delta}}{2}
\label{cubic1sol}
\end{eqnarray}
in terms of the discriminant of the corresponding quadratic equation
\begin{eqnarray}
\Delta && = \left[ (J_1^2+J_3^2) + (G_1^2+G_3^2) \right]^2 - 4  ( J_1 G_3-G_1 G_3)^2
\nonumber \\
&& =  \left[ (J_1^2+J_3^2) - (G_1^2+G_3^2) \right]^2 + 4  ( J_1 G_1+J_3 G_3)^2
\label{delta}
\end{eqnarray}
One can then write more explicitly the pseudo-spins of Eq. \ref{pseudospin},
but it is more instructive at this point to compute instead all the quasi-Majorana fermions 
from the different approach described below.

\subsection{ Dynamics within the subspace of odd operators }

The space of odd operators of dimension ${\cal N}_{op}^{odd} = 2^{2N-2}=2^4= 16$
contains ${\cal N}_{op}^{oddA}  = 10$ operators of flavor $A$
and ${\cal N}_{op}^{oddB}  = 6$ operators of flavor $B$ as we now describe.

\subsubsection{ Odd Operators of flavor $A$ (sector $T=+1$  of dimension $10$ )}

The basis $A_{\alpha}$ with $\alpha=1,..,10$ contains
the three operators involving a single Majorana fermion of flavor $a$
\begin{eqnarray}
A_1 = a_1
\nonumber    \\
A_2= a_2
\nonumber   \\
A_3= a_3
\label{a13}
\end{eqnarray}
the three operators involving $b_1$ and two Majorana fermion of flavor $a$
\begin{eqnarray}
A_4 = i b_1 a_2 a_3
\nonumber \\
A_5 =i b_1 a_1  a_3 
\nonumber \\
A_6=i  b_1 a_1 a_2
\label{a46}
\end{eqnarray}
 the three operators involving $b_2$ and two Majorana fermion of flavor $a$
\begin{eqnarray}
A_7 = i b_2 a_2 a_3
\nonumber \\
A_8 =i b_2 a_1  a_3 
\nonumber \\
A_9=i  b_2 a_1 a_2
\label{a79}
\end{eqnarray}
and finally the operator of Eq. \ref{upstot}
\begin{eqnarray}
A_{10} = \Upsilon^{tot}  = -a_1 b_1 a_2 b_2 a_3
\label{a10}
\end{eqnarray}

\subsubsection{ Odd Operators of flavor $B$ (sector $T=-1$  of dimension $6$ )}

The basis $B_{\beta}$ with $\beta=1,..,6$ contains
the two operators involving a single Majorana fermion of flavor $b$
\begin{eqnarray}
B_1 = b_1
\nonumber    \\
B_2=b_2 
\label{b12}
\end{eqnarray}
the operator involving the three Majorana fermions of flavor $a$
\begin{eqnarray}
B_3=i a_1 a_2 a_3
\label{b3}
\end{eqnarray}
and the three operators involving $b_1 b_2$ and one  Majorana fermion of flavor $a$
\begin{eqnarray}
B_4=i  b_1 b_2 a_1
\nonumber \\
B_5=i  b_1  b_2 a_2
\nonumber \\
B_6 = i   b_1 b_2 a_3
\label{b456}
\end{eqnarray}

\subsubsection{ Matrix $M_{\beta \alpha}$ of dimension $6 \times 9$ }

Since $A_{10}=\Upsilon^{tot} $ is already known to be a zero-mode, the rectangular real matrix $M_{\beta \alpha}$ of Eq. \ref{mbetaalpha}
is actually of dimension $6 \times 9$ and reads in terms of the nine couplings of Eq. \ref{h5}
\begin{eqnarray}
M =
 \begin{pmatrix}
I_1 & I_2 & I_3 & 0 & 0 & 0 &  -G_1 & G_2 & -G_3 \\
J_1 & J_2 & J_3 & G_1 & -G_2 & G_3 &  0 & 0 & 0 \\
0 & 0 & 0 & -I_1 & I_2 &-I_3 &  -J_1 & J_2 & -J_3 \\
0 & -G_3 & G_2 & 0 & -J_3 & - J_2 & 0 & I_3 & I_2 \\
G_3 & 0 & -G_1 & -J_3 & 0 & J_1 &  I_3 & 0 & - I_1 \\
-G_2 & G_1 & 0 & J_2 & J_1 & 0 &  -I_2 & -I_1 & 0 \\
\end{pmatrix}
\label{m69}
\end{eqnarray}
The corresponding symmetric square matrix $N=MM^{\dagger}$ of Eq. \ref{nmatrix} of dimension $6 \times 6$ reads
\begin{eqnarray}
N = 
 \begin{pmatrix}
\vert \vec I \vert^2+\vert \vec G \vert^2  & \vec I. \vec J & \vec J. \vec G & 2 (G_2 I_3-G_3I_2) & 2 (I_1 G_3-I_3 G_1) & 2(G_1 I_2- G_2 I_1) \\
\vec I. \vec J & \vert \vec J \vert^2+\vert \vec G \vert^2  & - \vec I. \vec G & 2 (G_2 J_3-G_3J_2)  & 2 (J_1 G_3-J_3 G_1) & 2(G_1 J_2- G_2 J_1)   \\
\vec J. \vec G  & -\vec I. \vec G  & \vert \vec I \vert^2+\vert \vec J \vert^2 &  2 (J_2 I_3-J_3I_2) &2 (I_1 J_3-I_3 J_1)  & 2(J_1 I_2- J_2 I_1)  \\
2 (G_2 I_3-G_3I_2) & 2 (G_2 J_3-G_3J_2) & 2 (J_2 I_3-J_3I_2)   & \vert \vec 2 \vert^2+\vert \vec 3 \vert^2 & -\vec 1. \vec 2 & - \vec 1. \vec 3  \\
2 (I_1 G_3-I_3 G_1) & 2 (J_1 G_3-J_3 G_1) & 2 (I_1 J_3-I_3 J_1)  & -\vec 1. \vec 2 & \vert \vec 1 \vert^2+\vert \vec 3 \vert^2 & - \vec 2. \vec 3  \\
2(G_1 I_2- G_2 I_1)  & 2(G_1 J_2- G_2 J_1) & 2(J_1 I_2- J_2 I_1)  & - \vec 1. \vec 3  & - \vec 2. \vec 3 & \vert \vec 1 \vert^2+\vert \vec 2 \vert^2  \\
\end{pmatrix}
\nonumber 
\end{eqnarray}
where we have introduced the notations $\vec 1=(I_1,J_1,G_1)$, $\vec 2=(I_2,J_2,G_2)$ and $\vec 3=(I_3,J_3,G_3)$ to simplify the expression of some matrix elements.

In the diagonalized form for pseudo-Majorana fermions corresponding to Eq \ref{hpseudoc},
the transformed matrix should become
\begin{eqnarray}
{\tilde N}=
 \begin{pmatrix}
\omega_1^2+\omega_3^2  & 0 & 0 & 0 & 0 & 2 \omega_1\omega_3 \\
0 &  \omega_2^2+\omega_3^2  & 0 & 0 & -2 \omega_2\omega_3  & 0   \\
0  & 0  & \omega_1^2+\omega_2^2 &  -2 \omega_1 \omega_2  & 0 & 0 \\
0 & 0 &  -2 \omega_1\omega_2  &\omega_1^2+\omega_2^2  & 0 & 0  \\
0 & -2 \omega_2\omega_3 & 0  & 0 &\omega_2^2+\omega_3^2  & 0  \\
2 \omega_1\omega_3 & 0 & 0 & 0  & 0 &\omega_1^2+\omega_3^2    \\
\end{pmatrix}
\label{ntilde}
\end{eqnarray}

To make some progress, it is now useful to analyze the possibilities for the expansions  in the basis $B_{\beta}$
of the two pseudo-Majorana operators of flavor $B$
\begin{eqnarray}
\tilde b_1 = \sum_{\beta=1}^6 y_{1 \beta} B_{\beta}
\nonumber \\
\tilde b_2 = \sum_{\beta=1}^6 y_{2 \beta} B_{\beta}
\label{tildeb12}
\end{eqnarray}
that should satisfy the anticommutation relations (Eq \ref{anticomm})
\begin{eqnarray}
 2 \delta_{ij} = \{\tilde b_i, \tilde b_j \} =\sum_{\beta=1}^6 y_{i \beta} \sum_{\beta'=1}^6 y_{j \beta'} \{   B_{\beta} ,   B_{\beta'} \} 
\label{tildeb12anti}
\end{eqnarray}
From the anticommutation properties of the operators $B_{\beta}$ for $\beta=1,..,6$,
one concludes that the only possibility is actually only a three-dimensional rotation $R_B$ (satisfying $R_B R_B^t =\mathbb{1} $)
in the subspace of the three first operators $(B_1=b_1,B_2=b_2,B_3=i a_1 a_2 a_3)$
that anticommute with each other
\begin{eqnarray}
 \begin{pmatrix}
\tilde b_1   \\
\tilde b_2  \\
\tilde B_3  &  \\
\end{pmatrix}
= R_B 
 \begin{pmatrix}
 b_1   \\
 b_2  \\
 B_3  &  \\
\end{pmatrix}
\label{brot}
\end{eqnarray}
Since we wish to put the matrix $N$ into the form of $\tilde N$ of Eq. \ref{ntilde}, $R_B$ should be chosen as the rotation that diagonalizes
the first $3 \times 3$ block of the matrix $N$
\begin{eqnarray}
N_{ (3 \times 3)} \equiv
 \begin{pmatrix}
\vert \vec I \vert^2+\vert \vec G \vert^2  & \vec I. \vec J & \vec J. \vec G \\
\vec I. \vec J & \vert \vec J \vert^2+\vert \vec G \vert^2  & - \vec I. \vec G    \\
\vec J. \vec G  & -\vec I. \vec G  & \vert \vec I \vert^2+\vert \vec J \vert^2   \\
\end{pmatrix}
= R_B^t \begin{pmatrix}
\omega_1^2+\omega_3^2  & 0 & 0  \\
0 &  \omega_2^2+\omega_3^2  & 0    \\
0  & 0  & \omega_1^2+\omega_2^2  \\
\end{pmatrix}  
R_B
\label{diagon}
\end{eqnarray}
If one focuses of the three eigenvalues only, one recovers 
via the characteristic polynomial that the three squares $(\omega_1^2,\omega_2^2,\omega_3^2)$
are the three roots of the cubic equation of Eq. \ref{cubic}.
The three-dimensional rotation $R_B$ depending on three Euler angles 
should be then computed from the corresponding eigenvectors.

To obtain simpler results for this rotation, it is convenient to focus on the particular case of Eq. \ref{special5},
where the roots of the cubic equation have been given in Eq. \ref{cubic1} and \ref{cubic1sol}.
The diagonalization problem of Eq. \ref{diagon}
\begin{eqnarray}
N_{ (3 \times 3)} && \equiv
 \begin{pmatrix}
I_2^2+(G_1^2+G_3^2)  & 0  & (J_1G_1+J_3G_3) \\
0 & (J_1^2+J_3^2)  +(G_1^2+G_3^2)    & 0    \\
  (J_1G_1+J_3G_3)  & 0  &  I_2^2+(J_1^2+J_3^2)   \\
\end{pmatrix}
\nonumber \\
&& = R_B^t \begin{pmatrix}
I_2^2+\omega_3^2  & 0 & 0  \\
0 &   (J_1^2+J_3^2)  +(G_1^2+G_3^2)  & 0    \\
0  & 0  & I_2^2+\omega_2^2  \\
\end{pmatrix}  
R_B
\label{diagonspecial}
\end{eqnarray}
now only involves a two-dimensional rotation concerning $(B_1,B_3)$ while $B_2$ is left unchanged
\begin{eqnarray}
R_B\equiv
 \begin{pmatrix}
\cos \theta  & 0  & - \sin \theta \\
0 & 1   & 0    \\
\sin \theta & 0  &  \cos \theta    \\
\end{pmatrix}
\label{rb2}
\end{eqnarray}
where the angle $\theta$ is determined by
\begin{eqnarray}
\cos(2 \theta) = \frac{\left[ (J_1^2+J_3^2) - (G_1^2+G_3^2) \right]}{ \sqrt{  \left[ (J_1^2+J_3^2) - (G_1^2+G_3^2) \right]^2 + 4  ( J_1 G_1+J_3 G_3)^2 } }
\nonumber \\
\sin(2 \theta) = \frac{ 2( J_1 G_1+J_3 G_3) }{ \sqrt{ \left[ (J_1^2+J_3^2) - (G_1^2+G_3^2) \right]^2 + 4  ( J_1 G_1+J_3 G_3)^2 } }
\label{2theta}
\end{eqnarray}

Now that the two pseudo-Majorana fermions of flavor $B$ have been determined,
it is convenient to recast this rotation $R_B$ into the unitary transformation 
\begin{eqnarray}
 U_B =e^{  \frac{\theta}{2} b_1 B_3 }  =  e^{ i \frac{\theta}{2} b_1 a_1 a_2 a_3 } = \cos \frac{\theta}{2}  + i \sin \frac{\theta}{2}  b_1 a_1 a_2 a_3
\label{ub}
\end{eqnarray}
that implement the above transformation 
\begin{eqnarray}
\tilde b_1 && = U_B b_1 U_B^{\dagger} = \cos \theta \  b_1 - i \sin \theta  \ a_1 a_2 a_3
\nonumber \\
\tilde b_2 && = U_B b_2 U_B^{\dagger} = b_2
\label{tildeb12u}
\end{eqnarray}
so that the corresponding transformation for the Majorana fermions of flavor $a$ needed to maintain the anticommutation relations reads
\begin{eqnarray}
\tilde a_1 && = U_B a_1 U_B^{\dagger} = \cos \theta \  a_1 + i \sin \theta  \ b_1 a_2 a_3
\nonumber \\
\tilde a_2 && = U_B a_2 U_B^{\dagger} = \cos \theta \  a_2 - i \sin \theta  \ b_1 a_1 a_3
\nonumber \\
\tilde a_3 && = U_B a_3 U_B^{\dagger} = \cos \theta \  a_3 + i \sin \theta  \ b_1 a_1 a_2
\label{tildea123u}
\end{eqnarray}

In terms of these new Majorana operators, the Hamiltonian becomes
\begin{eqnarray}
H  && = i I_2 b_1  a_2 +  i b_2 (J_1 a_1 +J_3 a_3) + b_1b_2 ( G_1 a_2 a_3 + G_3 a_1 a_2)
\nonumber \\
&& =i I_2 \tilde b_1  \tilde a_2
+ i  \tilde b_2 \left(  (J_1 \cos \theta+ G_1 \sin \theta ) \tilde  a_1 + (J_3 \cos \theta+ G_3 \sin \theta )  \tilde  a_3  \right)
\nonumber \\
&& + \tilde b_1\tilde b_2 \left(  (G_1 \cos \theta- J_1 \sin \theta ) \tilde  a_2 \tilde  a_3  + (G_3 \cos \theta- J_3 \sin \theta )  \tilde  a_1   \tilde  a_2 \right)
\label{h5inter}
\end{eqnarray}

We know that $(\tilde b_1,\tilde b_2)$ are the appropriate pseudo-Majorana fermions of flavor $B$ that will diagonalize $H$,
while in the sector of flavor $A$, the operators $(\tilde a_1,\tilde a_2,\tilde a_3)$ of Eq. \ref{tildea123u}
have only been determined to respect the anticommutation relations, so one still needs to perform a rotation within the sector $A$
to obtain the quasi-Majorana operators $ (\tilde{\tilde a}_1,\tilde{\tilde a}_2,\tilde{\tilde a}_3)$ that diagonalize $H$. 
From Eq. \ref{h5inter}, one obtains that the operator coupled to $\tilde b_1 $ is $\tilde a_2 $, so that this operator should be kept
\begin{eqnarray}
\tilde{\tilde a}_2 =  \tilde a_2
\label{keep2}
\end{eqnarray}
while the operator coupled to $\tilde b_2 $ is a linear combination of $ (\tilde{\tilde a}_1,\tilde{\tilde a}_3)$
that should define the appropriate new operator  $\tilde{\tilde a}_3$, so one needs to perform the following rotation
\begin{eqnarray}
\tilde{\tilde a}_1  && =\cos \phi  \ \tilde  a_1 - \sin \phi \  \tilde  a_3  
\nonumber \\
\tilde{\tilde a}_3  && =\sin \phi  \ \tilde  a_1 + \cos \phi \  \tilde  a_3  
\label{a3tt}
\end{eqnarray}
of angle $\phi$ satisfying
\begin{eqnarray}
\cos \phi  && = \frac{   (J_3 \cos \theta+ G_3 \sin \theta )  }
{\sqrt{  (J_1 \cos \theta+ G_1 \sin \theta )^2 + (J_3 \cos \theta+ G_3 \sin \theta )^2  } } 
\nonumber \\
\sin \phi  && = \frac{  (J_1 \cos \theta+ G_1 \sin \theta ) }
{\sqrt{  (J_1 \cos \theta+ G_1 \sin \theta )^2 + (J_3 \cos \theta+ G_3 \sin \theta )^2 } } 
\label{phi}
\end{eqnarray}
The square of the denominator reads using Eqs \ref{2theta} defining the angle $\theta$
\begin{eqnarray}
&& (J_1 \cos \theta+ G_1 \sin \theta )^2 + (J_3 \cos \theta+ G_3 \sin \theta )^2
 = (J_1^2+J_3^2) \cos^2 \theta +  (G_1^2+G_3^2) \sin^2 \theta +2 (J_1 G_1+J_3 G_3) \cos \theta \sin \theta
\nonumber \\
&& = (J_1^2+J_3^2) \frac{1+\cos(2 \theta)}{2}  +  (G_1^2+G_3^2) \frac{1-\cos(2 \theta)}{2} + (J_1 G_1+J_3 G_3)  \sin (2 \theta)
\nonumber \\
&& =\frac{(J_1^2+J_3^2) +(G_1^2+G_3^2)  }{2}  
+   \frac{\left[ (J_1^2+J_3^2) - (G_1^2+G_3^2) \right]^24 ( J_1 G_1+J_3 G_3)^2}{ 2 \sqrt{  \left[ (J_1^2+J_3^2) - (G_1^2+G_3^2) \right]^2 + 4  ( J_1 G_1+J_3 G_3)^2 } }
\nonumber \\
&& =\frac{(J_1^2+J_3^2) +(G_1^2+G_3^2) +  \sqrt{  \left[ (J_1^2+J_3^2) - (G_1^2+G_3^2) \right]^2 + 4  ( J_1 G_1+J_3 G_3)^2 }  }{2}  
= \omega_2^2
\label{deno2}
\end{eqnarray}
i.e. it coincides with $\omega_2^2$ of Eq. \ref{cubic1sol} as it should for consistency.
In terms of the operators of Eq. \ref{a3tt}, Eq. \ref{h5inter} thus becomes, using also $\omega_1=I_2$ and $\omega_2=\sqrt{\omega_2^2}$
\begin{eqnarray}
H && =i \omega_1 \tilde b_1  \tilde a_2
+ i  \omega_2 \tilde b_2 \tilde{\tilde a}_3
+ \tilde b_1\tilde b_2 ( \tilde{\tilde G}_1  \tilde  a_2 \tilde{\tilde a}_3 + \tilde{\tilde G}_3  \tilde{\tilde a}_1 \tilde  a_2 )
\label{h5interbis}
\end{eqnarray}
where the new interaction couplings
\begin{eqnarray}
\tilde{\tilde G}_1 && = (G_1 \cos \theta- J_1 \sin \theta )   \cos \phi 
 - (G_3 \cos \theta- J_3 \sin \theta )   \sin \phi   
\nonumber \\
\tilde{\tilde G}_3 && =  (G_1 \cos \theta- J_1 \sin \theta )  \sin \phi  
 + (G_3 \cos \theta- J_3 \sin \theta )  \cos \phi   
\label{gtildeinter}
\end{eqnarray}
can be simplified using Eq \ref{phi} and \ref{deno2} for the angle $\phi$ 
and Eqs \ref{2theta} for the angle $\theta$ to obtain
\begin{eqnarray}
\tilde{\tilde G}_1 && = (G_1 \cos \theta- J_1 \sin \theta )   \frac{   (J_3 \cos \theta+ G_3 \sin \theta )  }{\omega_2} 
 - (G_3 \cos \theta- J_3 \sin \theta )   \frac{  (J_1 \cos \theta+ G_1 \sin \theta ) }{\omega_2}  
 = \frac{ G_1J_3-G_3 J_1 } { \omega_2} 
\nonumber \\
\tilde{\tilde G}_3 && =  (G_1 \cos \theta- J_1 \sin \theta )  \frac{  (J_1 \cos \theta+ G_1 \sin \theta ) }{\omega_2}  
 + (G_3 \cos \theta- J_3 \sin \theta )   \frac{   (J_3 \cos \theta+ G_3 \sin \theta )  }{\omega_2}  
\nonumber \\
&& = \frac{ 2 (J_1G_1+J_3G_3) \cos(2\theta) - (J_1^2+J_3^2-G_1^2-G_3^2 ) \sin(2 \theta) } {2 \omega_2} =0
\label{gtilde}
\end{eqnarray}
In conclusion, $\tilde{\tilde G}_3$ vanishes as it should to make Eq. \ref{h5interbis} the diagonalized form of the Hamiltonian,
while the square of $\tilde{\tilde G}_1$ coincides with the third root $\omega_3^2$ ( Eq. \ref{cubic23})
\begin{eqnarray}
\tilde{\tilde G}_1^2 && = \frac{ (G_1J_3-G_3 J_1)^2 } { \omega_2^2} =\omega_3^2
\label{gtildecarre}
\end{eqnarray}
as it should for consistency.

In summary, the Hamiltonian of the special case is diagonalized in the pseudo-Majorana fermions
\begin{eqnarray}
\tilde{\tilde b}_1 && =\tilde b_1  =  \cos \theta \  b_1 - i \sin \theta  \ a_1 a_2 a_3
\nonumber \\
\tilde{\tilde b}_2 && = \tilde b_2 = b_2
\nonumber \\
\tilde{\tilde a}_2 && =  \tilde a_2 = \cos \theta \  a_2 - i \sin \theta  \ b_1 a_1 a_3
\nonumber \\
\tilde{\tilde a}_1  && =\cos \phi  \ \tilde  a_1 - \sin \phi \  \tilde  a_3  
= \cos \phi  (\cos \theta \  a_1 + i \sin \theta  \ b_1 a_2 a_3 ) - \sin \phi (\cos \theta \  a_3 + i \sin \theta  \ b_1 a_1 a_2) 
\nonumber \\
\tilde{\tilde a}_3  && =\sin \phi  \ \tilde  a_1 + \cos \phi \  \tilde  a_3  
= \sin \phi  (\cos \theta \  a_1 + i \sin \theta  \ b_1 a_2 a_3 ) + \cos \phi (\cos \theta \  a_3 + i \sin \theta  \ b_1 a_1 a_2)
\label{tildetildefinal}
\end{eqnarray}

While we have chosen here to focus on the special case of Eq. \ref{special5} to obtain more explicit results,
it seems now useful to point out what changes are needed to obtain the pseudo-Majorana fermions for the general case of Eq \ref{h5} :

(i) the three-dimensional rotation $R_B$ that diagonalizes Eq \ref{diagon} will involves three Euler angles (instead of the single angle $\theta$ of Eq. \ref{rb2})
so that the unitary transformation of Eq. \ref{ub} will now contain three angles, for instance one could choose the parametrization
\begin{eqnarray}
 U_B =  e^{  \frac{\theta_1}{2} b_2 B_3 }  e^{  \frac{\theta_2}{2} b_1 B_3 }  e^{  \frac{\theta_3}{2} b_1 b_2 }  
\label{ub3angles}
\end{eqnarray}

(ii) the rotation from $(\tilde a_1,\tilde a_2,\tilde a_3)$ to $(\tilde{\tilde a}_1,\tilde{\tilde a}_2,\tilde{\tilde a}_3 )$
will also be a general three-dimensional rotation involving three Euler angles $(\phi_1,\phi_2,\phi_3)$  (instead of the single angle $\phi$ of Eq \ref{a3tt})
\begin{eqnarray}
 U_A =  e^{  \frac{\phi_1}{2} \tilde a_2\tilde a_3 }  e^{  \frac{\phi_2}{2} \tilde a_1\tilde a_3 }    e^{  \frac{\phi_3}{2} \tilde a_1\tilde a_2 }   
\label{ua3angles}
\end{eqnarray}

The three Euler angles $\theta_{1,2,3}$, the three Euler angles $\phi_{1,2,3}$ and the three pseudo-couplings $\omega_{1,2,3}$
correspond to the nine parameters that are needed to diagonalize the general Hamiltonian of Eq \ref{h5} containing nine couplings.

\section{ Conclusion }

\label{sec_conclusion}

For random interacting Majorana models where the only symmetries are the Parity $P$ and the Time-Reversal-Symmetry $T$,
we have compared various approaches to construct exact even and odd normalized zero modes in finite size,
with explicit examples for small systems. 

For even normalized zero modes known as the commuting pseudo-spins $\tau^z_j$ that diagonalize the Hamiltonian,
we have described how the Yang and Feldman idea to consider the powers of the Hamiltonian \cite{feldman} 
could be used to construct directly the pseudo-spins $\tau^z_j$ and their pseudo-couplings $\omega_{j_1,j_2..}$,
without computing first the many-body-eigenstates.

For odd normalized zero modes, we have adapted the Goldstein and Chamon approach concerning an odd number of Majorana fermions \cite{goldstein}
to the presence of the Time-Reversal symmetry $T$, where the orthonormal basis $\Upsilon_{\mu}$ of the odd operators subspace
can be decomposed into operators $A_{\alpha}$ and $B_{\beta}$ of flavor $A$ and $B$ respectively.
We have explained how the Goldstein-Chamon matrix ${\cal H}_{\mu\nu}$ can be then reshaped into a rectangular real matrix $M_{\beta \alpha}$,
and how the quasi-Majorana fermions that diagonalize the Hamiltonian could be then computed.

The practical application of these methods to large systems clearly goes beyond the scope of the present paper
and certainly requires a much better understanding of the structure of the space of odd operators 
with respect to commutation/anticommutation :
indeed on the example concerning five Majorana operators, we have seen that the anticommutation relations 
significantly restrict the form of the pseudo-operators of flavor $B$ (from Eq \ref{tildeb12} to Eq. \ref{brot}),
so that the appropriate generalization of this simplification to systems of arbitrary size seems an important issue
that requires further work.

However from a real-space renormalization perspective,
the explicit construction of pseudo-spins and pseudo-Majorana operators is actually very interesting already for small systems, 
where they not only allow to re-interpret previous block-spin procedures (see section VI E)
but they also suggest new procedures based on blocks containing odd numbers of Majorana fermions (see section VI F).

In conclusion, the idea to replace the diagonalization of the Hamiltonian H in terms of the $2^N$ many-body eigenstates
by the diagonalization in terms of $N$ pseudo-spins $\tau^z_j$ or $(2N)$ pseudo-Majorana fermions $\tilde \gamma_j$
can be considered as one of the most interesting that has emerged recently,
 with far-reaching consequences that have yet to be fully understood, both in general and in 
the various specific contexts mentioned in the Introduction :

(i) in the field of Many-Body-Localization, the $N$ pseudo-spins are expected to be spatially localized in the Many-Body-Localized phase, but an essential issue is the characterization of their spatial properties at the MBL-phase-transition
towards the delocalized ergodic phase.

(ii) in the field of pure models displaying edge Majorana zero modes, such as the integrable XYZ chain studied in 
Ref \cite{strongzeromode}, it would be interesting to write the other quasi-Majorana fermions,
and to elucidate what are the simplifications induced by quantum integrability with respect to the general structure
of Eq. \ref{hmbl} valid for random models.

\appendix

\section{ Translation in the quantum spin chain language }

\label{sec_app}

For a chain of $N$ quantum spins described by Pauli matrices, 
the $(2N)$ string operators 
\begin{eqnarray}
a_j && \equiv \left( \prod_{k=1}^{j-1} \sigma_k^z \right) \sigma_j^x
\nonumber \\
b_j && \equiv  \left( \prod_{k=1}^{j-1} \sigma_k^z \right)  \sigma_j^y 
\label{sigmaxy}
\end{eqnarray}
 satisfy  the Majorana anticommutation relations of Eq \ref{anticomm} with the identification $\gamma_{2j-1}=a_j$ and $\gamma_{2j}=b_j$.

The parity of Eq. \ref{paritytotal} becomes
\begin{eqnarray}
P= \prod_{j=1}^N (-\sigma_j^z)
\label{paritysigmaz}
\end{eqnarray}

The Time-Reversal Symmetry $T$ of Eq. \ref{time} acts on the Pauli matrices as
\begin{eqnarray}
T  \sigma_j^x T^{-1} && =   \sigma_j^x
\nonumber \\
T   \sigma_j^y T^{-1} &&=  -  \sigma_j^y
\nonumber \\
T   \sigma_j^z T^{-1} && =  \sigma_j^z
\label{timespins}
\end{eqnarray}
Why this choice is possible for quantum spins is explained in the Lecture Notes \cite{lecture}, where many other subtleties of
the Time-Reversal Symmetry $T$ in quantum mechanics are discussed in detail.

When the Hamiltonian for the $(2N)$ Majorana fermions $(\gamma_1,...,\gamma_{2N})$
 does not involve the last one $\gamma_{2N}$ (Eq. \ref{hgn}), this means in the spin language that the Hamiltonian
does not involve the operators $\sigma_N^y$ and $\sigma_N^z$, so $H$ depends on the last quantum spin $\sigma_N$
only via the operator $\sigma_N^x$ so that they commute
\begin{eqnarray}
 [H ,\sigma_N^x ]=0
\label{hgns}
\end{eqnarray}
The operator ${\Upsilon}^{tot} $ of Eq. \ref{upstot} becomes in the spin language
\begin{eqnarray}
{\Upsilon}^{tot} =  i^{N-1}   \gamma_{1} \gamma_{2} ... \gamma_{2N-2}\gamma_{2N-1} =  (-1)^{N-1} \sigma_N^x
\label{upstotspin}
\end{eqnarray}
and is the basic normalized odd zero mode that relates the two states of the same energy in the two Parity sectors  $P=\pm 1$ (Eq. \ref{none}).

In section \ref{sec_three}, the Hamiltonian of Eq. \ref{h3} respecting $P$ and $T$ and depending only on the three first Majorana fermions becomes in the spin language
the two first terms of the quantum Ising chain
\begin{eqnarray}
H=  -K_1 \sigma^z_1- K_2 \sigma_1^x\sigma_2^x
\label{h3spin}
\end{eqnarray}

In section \ref{sec_five},
the Hamiltonian of Eq \ref{h5} respecting $P$ and $T$ and depending only on the first five Majorana fermions translates in the spin language into
\begin{eqnarray}
H_5  && = i b_1 (I_1 a_1+I_2 a_2 +I_3 a_3) +  i b_2(J_1 a_1+J_2 a_2 +J_3 a_3)   + b_1b_2 ( G_1 a_2 a_3- G_2 a_1 a_3 + G_3 a_1 a_2)
\nonumber \\
&& = I_1 \sigma^z_1- I_2 \sigma_1^x\sigma_2^x- I_3 \sigma_1^x \sigma_2^z \sigma_3^x
- J_1 \sigma_1^y \sigma^y_2 +J_2  \sigma^z_2 - J_3 \sigma_2^x \sigma^x_3
+ G_1 \sigma^x_1 \sigma^x_3 +G_2 \sigma^z_1 \sigma_2^x \sigma^x_3+ G_3 \sigma^z_1 \sigma^z_2
\label{h5spin}
\end{eqnarray}
while the special case of Eq. \ref{special5} involves only five couplings
\begin{eqnarray}
H_5^{special}   = - I_2 \sigma_1^x\sigma_2^x
- J_1 \sigma_1^y \sigma^y_2  - J_3 \sigma_2^x \sigma^x_3
+ G_1 \sigma^x_1 \sigma^x_3 + G_3 \sigma^z_1 \sigma^z_2
\label{h5spinspecial}
\end{eqnarray}

\end{document}